# Auto-compensating differential phase shift quantum key distribution

**Xiaohong Han, Guang Wu, Chunyuan Zhou, and Heping Zeng[1]**

Key Laboratory of Optical and Magnetic Resonance Spectroscopy, and Department of Physics,

East China Normal University, Shanghai 200062, People's Republic of China

Abstract

We propose an auto-compensating differential phase shift scheme for quantum key distribution with a high key-creation efficiency, which skillfully makes use of automatic alignment of the photon polarization states in optical fiber with modified Michelson interferometers composed of unequal arms with Faraday mirrors at the ends. The Faraday-mirrors-based Michelson interferometers not only function as pulse splitters, but also enable inherent compensation of polarization mode dispersion in the optic-fiber paths at both Alice's and Bob's sites. The sequential pulses encoded by differential phase shifts pass through the quantum channel with the same polarization states, resulting in a stable key distribution immune to the polarization mode dispersion in the quantum channel. Such a system features perfect stability and higher key creation efficiency over traditional schemes.

---

[1]To whom correspondence should be addressed. Electronic address: hpzeng@phy.ecnu.edu.cn.

After the creative work of Brassard and Bennett [1,2], scientists have done many encouraging experiments for quantum key distribution (QKD) [3-12]. Due to its promising prospect in crypto-communication, people are mostly attracted by the work done in optical fiber [4-12], which were roughly classified into two categories as "one way" and "plug and play" (P&P) schemes. The first category is represented by Bennett's proposal of a double Mach-Zehnder system [2]. The main difficulty associated with this QKD scheme lies in two aspects. Firstly, the polarization mode dispersion (PMD) introduced by imbalanced arms causes variable degradation of photon interferences. Secondly, the arm differences of Alice's and Bob's interferometers must be kept stable within a fraction of the photon wavelength to maintain correct phase relations. To avoid inconvenience of active control of phase drifts and PMD, Muller et al. [4] invented a round-way "P&P" QKD system based on polarization-orthogonal reflection from Faraday mirrors (FMs), which enables high-quality interference without any initial calibration step or active compensation in the optical circuits. With the P&P scheme, Zbinden et al. successfully implemented the 67 km out-door QKD experiment [5]. But this scheme still bears some deficiency. For example, the requirement that the signal travel both directions along the transmission line not only leads to nontrivial technical difficulties such as Rayleigh backward scattering of the strong forward pulses, but may even leave a back door for Eve to make "Trojan horse" attack. On the other hand, a round-way scheme is not preferred if one considers to use ideal single photons as information carriers. From the practical points of view, a QKD system based on either one-way or round-way scheme faces a troublesome problem of impractically low rate of key distribution, which is mainly limited by the immaturity of up-to-date techniques for single-photon generation and detection. To enhance the secret key creation rate, Yamamoto et al proposed a differential phase shift (DPS) scheme by splitting a single-photon pulse into three sequential time slots, followed by DPS encoding with half-wave phase modulations randomly applied to the adjacent components of the photon. At the detection stage, the key creation efficiency was as high as 2/3 [6].

In this letter, we demonstrate that the virtue points of PMD auto-compensation from the FM-based design and high key-creation efficiency from DPS encoding can be combined on a one-way optical-fiber structure that skillfully uses FM-armed Michelson interferometers (FM-MI), where the FM-MIs consist of unequal arms tailed with FM reflections. At Alice's station, the FM-MI allows an automatic alignment of the polarization states of the sequential DPS pulses to pass through the long-distance quantum channel with the same polarization states, reaching Bob's station with the same unknown polarization states. At Bob's station, any possible polarization difference between the interfering pulses due to PMD in the unequal arms is also auto-compensated by the FM-MI. This design eventually results in a stable key distribution immune to the PMD in the quantum channel, typically a long-distance optical fiber. Such a DPS QKD scheme features perfect stability, high key creation efficiency, immunity to "Trojan horse" attack, and capability to be used for ideal single photons.

Figure 1 shows the setup of the proposed auto-compensating DPS QKD scheme. Alice uses a FM-MI composed of a 2×3 coupler (CP1) with two ports on one side and three on the other which are connected with three unequal arms tailed with three FMs (FM1-FM3), while Bob uses a FM-MI which consists of a 2×2 coupler (CP2) and two FMs (FM4 and FM5). In Alice's site, a photon from the single-photon generator is divided into three paths (a, b, and c) with the same length differences between sequential paths by CP1 which makes the probabilities for a photon to pass through each route equally. There is an FM at the end of each path, which reflects the input pulse back to the 2×3 coupler (CP1), and rotates the input polarization to its orthogonal state. By using FMs, the system automatically compensates for any birefringence effects and polarization dependent losses in Alice's site. The time delays between paths a and b and between b and c are equally fixed as T. The phase modulator in Alice's site gives the recombined photon a random modulation by 0 or $\pi$ to each of the three time slots [6]. Bob uses his FM-MI and single-photon detectors to "read out" the encoded quantum keys as follows. The incoming single-photon pulses are directed into two paths with the same probabilities by CP2, and recombined by the same coupler after reflection by FMs at the ends of both arms. The polarizations of the reflected photons become orthogonal to the incoming ones, and the time delay between the long path and the short path is also T. As illustrated in Fig. 1, there are four time instances counted at Bob's detectors and only two of them (t2 and t3) contain interferential information. The amplitudes of the interferential signals corresponding to phase differences between two sequential pulses, are detected at this two time instances, from which, Bob knows the phase information added to the pulses by Alice, i.e., he obtains the secret key. After this stage, Bob publicly announces the photon clicking time, so that Alice can figure out the Bob's detection results merely according to the time information and screen out the useful information as secret keys. Thus, Alice and Bob share the same bit string and an integrated communication is achieved. As pulses a and c contribute 50% while the pulse b contributes 100% to the key distribution, the total key creation efficiency is 2/3.

High efficiency is a strong point of the DPS scheme. In conventional QKD systems the most popular protocols employed are phase-encoding BB84 [1] and B92 [2]. It has been proved that the efficiency in a DPS scheme was much higher than that in conventional system and could be readily expanded by increasing the time slots [6, 11]. Our system is especially convenient to expand without making much change of the setup. To show this, two kinds of expandable systems are presented in Figs. 2 and 3. Fig. 2 (a) presents a scheme to split a photon into four time instances. In order to get more time slots, two FM-MIs are connected in series in Alice's site to replace the 2×3 FM-MI in Fig. 1. Two couplers and four FMs are needed to form the two cascaded FM-MIs. The first FM-MI divides the photon from a single photon source into two pulses with a time delay 2T and then the two pulses are respectively divided into two pulses with a time delay T by the second FM-MI, resulting in four sequential pulses (a, b, c, and d) with the same time interval T. The four pulses are randomly modulated with a phase 0 or π. In Bob's site, there are five time instances as shown in Fig. 2 (a). As pulses a and d contribute 50% while the pulses b and c contributes 100% to the key distribution, the key creation efficiency of this system is 3/4. Higher efficiency can be achieved by increasing the number of the FM-MIs connected in series as shown in Fig. 2 (b). If n FM-MIs are used in Alice's site, the key creation efficiency reaches up to $(2^n - 1)/2^n$. As shown in Fig. 3, FM-MIs in parallel can also be used to increase the efficiency up to $n/(n+1)$, where n means the number of the couplers used in Alice's site.

The security of DPS QKD has already been partly demonstrated in Refs. [6, 12]. The proposed system is also immune from Trojan horse attack though FM reflection is used. In a standard two-way quantum cryptography system, Eve may eavesdrop the secret key information by using an exploring light pulse to experience the same phase modulations in Alice's site as the single-photon pulse. In the proposed system, Alice can simply put an isolator to ensure unidirectional propagation of the single-photon pulses from Alice to Bob, obviating divulgence of the secret phase modulations to any possible exploring light pulses.

The present scheme is immune to PMD influence in the quantum channel. In a traditional system with a Mach-Zehnder interferometer, PMD in the imbalanced arms debases the stability of the interference visibility. Since FMs are used in our system for polarization-orthogonal reflection at the unbalanced arms of the FM-MIs [13], PMD for the pulses on their ways to FMs are automatically compensated on their return ways. While the pulses are traveling through the quantum channel, the sequential pulses, which enter the quantum channel in Alice's site with the same polarization states, reach Bob's site with the same polarization states even though PMD in the quantum channel makes random polarization changes. FMs are also used in Bob's site to guarantee that the interfering pulses have the same polarization state. Taken as an example, the changes of polarization state are shown in Fig. 1. As a consequence, the single-photon interference becomes independent upon the PMD of the quantum channel.

To further illustrate the immunity of polarization changes in the quantum channel for the present scheme, we briefly discuss in what follows the transformation of polarization in the system by using Jones Matrix [13]. Suppose $A_n$ represents the Jones Matrix of fiber in the n-th fiber-paths of the FM-MI in Alice's site, $Q$ the Jones matrix of the long-distance fiber (quantum channel), and $B_0$ and $B_T$ the Jones Matrices of fiber in the respective paths of the FM-MI in Bob's site. The total transformation matrices can be described by $P \propto [B_o^+ M B_0 e^{i\beta_0} + B_T^+ M B_T e^{i\beta_1}] \cdot Q e^{i\phi} \cdot \sum_n [A_n^+ M A_n e^{i\alpha_n + i\varphi_n}]$, where $M$ represents the Jones matrix of an FM, $\alpha_n$ and $\beta_n$ are the round-trip phases through the n-th fiber-path of Alice's and Bob's FM-MIs, $\varphi_n$ is the modulated phases for the n-th pulse through the PM, and $\phi$ is the phase through the transmission fiber. Single-photon interference takes place at the time slot nT with the corresponding output Jones vector of the electric field given by $E_{out} \propto \{B_0^+ MB_0 \cdot Q \cdot A_{n+1}^+ MA_{n+1} e^{i(\alpha_{n+1}+i\varphi_{n+1}+\phi+\beta_0)} + B_T^+ MB_T \cdot Q \cdot A_n^+ MA_n e^{i(\alpha_n+\varphi_n+\beta_1+\phi)}\} E_{in}$, where $E_{in}$ represents the Jones vector of the input field. The output power at the time slot nT is given by

$$P_{out} \propto 2P_{in} + E_{in}^+ \begin{bmatrix} A_{n+1}^+ MA_{n+1} \cdot Q^+ \cdot B_0^+ MB_0 \cdot B_T^+ MB_T \cdot Q \cdot A_n^+ MA_n e^{-i(\Delta\alpha+\Delta\beta+\Delta\varphi)} \\ + A_n^+ MA_n \cdot Q^+ \cdot B_T^+ MB_T \cdot B_0^+ MB_0 \cdot Q \cdot A_{n+1}^+ MA_{n+1} e^{i(\Delta\alpha+\Delta\beta+\Delta\varphi)} \end{bmatrix} E_{in}, \quad (1)$$

where $\Delta\alpha = \alpha_1 - \alpha_2, \Delta\beta = \beta_2 - \beta_1, \Delta\varphi = \varphi_1 - \varphi_2$. Note that $A_n$, $B_n$, and $Q$ are time-dependent due to the randomness of the PMD, the interferential terms depend on not only polarization changes in Alice's and Bob's sites but also PMD in the transmission fiber. If polarization-dependent loss is negligible, all the Jones matrices are unitary and the unique features of the Jones matrix of the FM [13] ensure $A_{n+1}^+ MA_{n+1} \cdot Q^+ \cdot B_0^+ MB_0 \cdot B_T^+ MB_T \cdot Q \cdot A_n^+ MA_n = I$ and $A_n^+ MA_n \cdot Q^+ \cdot B_T^+ MB_T \cdot B_0^+ MB_0 \cdot Q \cdot A_{n+1}^+ MA_{n+1} = I$. The output power can then be simplified as $P_{out} \propto P_{in}[1+\cos(\Delta\alpha+\Delta\beta+\Delta\varphi)]$, which means a complete immunity of interferential output power from the PMD influence in the quantum channel. Nevertheless, $P_{out}$ is still dependent on the phase drifts Δα and Δβ, which are not invariable due to arm-length changes of Alice's and Bob's FM-MIs. Typically, the phase drifts are slow and can be compensated by using a feedback control. For such a purpose, a servo-system can be used with an injected pulse to offer the feedback signal. The injected pulses propagate along the same route as the single-photon pulse, but it is behind the single-photon pulse with a time delay T0>2T, in Alice's site, we can add phase modulation to the single-photon pulses while keep the injected light to be non-modulated by Alice's phase modulator (PMA), which ensures that the injected light would not bring any insecurity to the scheme. The servo-system records the interferential signal of the injected light pulse to detect the slow phase drifts $\Delta\alpha$ and $\Delta\beta$, and then gives a feedback to adjust the DPS phase modulations in Alice's phase modulator (PMA) to ensure that $\Delta\alpha + \Delta\beta + \Delta\varphi$ is 0 or $\pi$. In this way, the stability of the scheme is guaranteed.

In summery, we have introduced an auto-compensating differential phase shift quantum key distribution scheme with the help of FM-based Michelson interferometers. The environment induced polarization drift was automatically compensated by the to-and-fro mechanism in the local sites while the phase drift was stabilized using a mature technique of servo-system. Expanded versions with higher key creation efficiencies were also presented.

This work was funded in part by National Key Project for Basic Research (Grant 2001CB309300), National Natural Science Fund (Grant 10374028), key project sponsored by National Education Ministry of China (Grant 104193), and Science and Technology commission of Shanghai (Grant 03XD14012).

Captions

Figure 1 Setup of the auto-compensating DPS QKD system and the polarizations of pulses at different parts of the system. In Alice's site, a single-photon generator (SPG) supplies single-photon pulses for information carriers, which are connected with a 2×3 coupler (CP1) after an isolator (ISO1). After Alice's FM-MI, the single-photon pulses are DPS encoded by using a phase modulator (PMA). In Bob's site, a circulator (CIR) is connected in series in the long-distance fiber (LF) and the third port of the circulator is connected to a single-photon detector (SPD1) and a standard photon detector (PIN) through a wavelength-division multiplexer (C2). Another single-photon detector (SPD2) joins one port of the 2×2 coupler (CP2) with a wavelength-division multiplexer (C3). A servo-system (SERVO) is used to control the arm length drifts of the FM-MIs by using a laser diode (LD) to provide an injected light source through a wavelength-division multiplexer (C1), where the injected light from LD is detected by a photodetector (PIN).

Figure 2 Setup of the auto-compensating DPS QKD system with FM-MIs in series for a key creation efficiency of 3/4(a) and 7/8(b).

Figure 3 Setup of the auto-compensating DPS QKD system with FM-MIs in parallel for a key creation efficiency of 3/4(a) and 7/8(b).

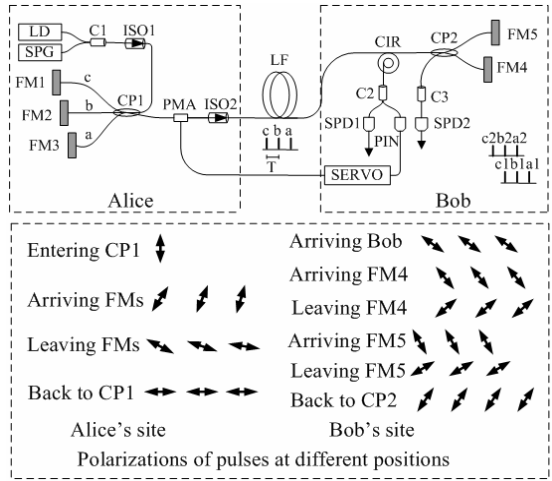

FIG.1

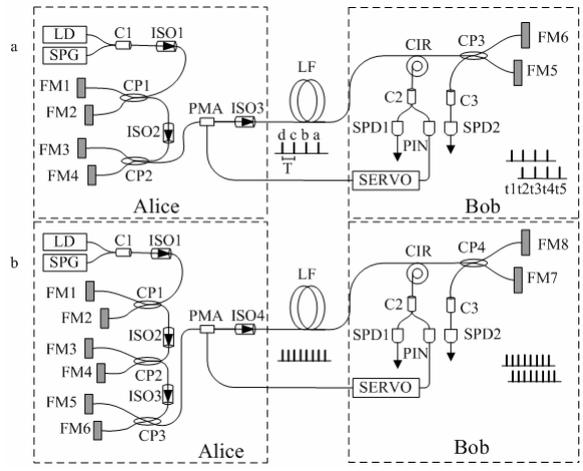

Fig.2

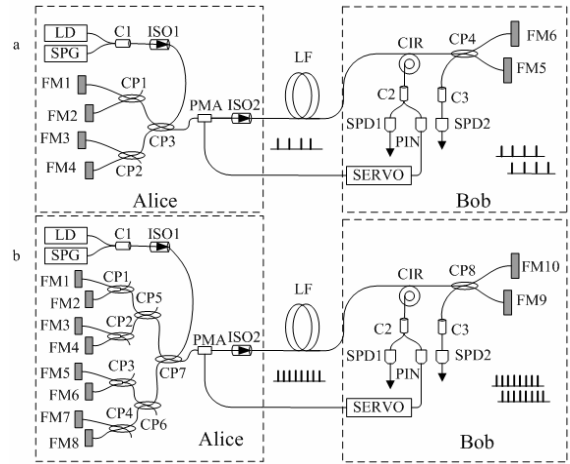

**Fig.3**